\documentclass[a4paper,11pt]{article}
\usepackage{pos}

\usepackage[utf8]{inputenc}
\usepackage{graphicx}
\usepackage{xcolor}
\usepackage{bm}
\usepackage{hyperref}
\hypersetup{colorlinks=true,linkbordercolor=blue,linkcolor=blue, citecolor=blue,pdfborderstyle={/S/U/W 1}}
\usepackage{xspace}

\def\be{\begin{equation}}
\def\ee{\end{equation}}
\def\bea{\begin{eqnarray}}
\def\eea{\end{eqnarray}}

\newcommand{\eq}[1]{Eq.~(\ref{#1})}
\newcommand{\jpsi}{\ensuremath{J/\psi}\xspace}

\newcommand{\dd}{{\rm d}}
\newcommand{\sqrts}{\sqrt{s}}
\newcommand{\pt}{\ensuremath{p_{_\perp}}}
\newcommand{\dsigpp}{\dd\sigma_{\rm pp}}

\newcommand{\RAA}{\ensuremath{R_{\textnormal{AA}}}\xspace}

\newcommand{\qhat}{\hat{q}}

\newcommand\meanepsbar{\ensuremath{\bar{\epsilon}}\xspace} 
\newcommand{\meaneps}{\ensuremath{\langle\epsilon\rangle}\xspace}
\newcommand{\TeV}{\ensuremath{\,\text{Te\hspace{-.08em}V}}\xspace}

\newcommand{\dNch}{\ensuremath{\frac{\dd N_{\rm ch}}{\dd y}}\xspace}
\newcommand{\dNg}{\ensuremath{\frac{\dd N_{k}}{\dd y}}\xspace}

\newcommand{\AT}{\ensuremath{A_{\perp}}\xspace}
\newcommand{\qzero}{\ensuremath{\hat{q}_0}\xspace}
\newcommand{\PbPb}{\ensuremath{\text{PbPb}}\xspace}

\newcommand{\ecc}{\ensuremath{\textnormal{e}}\xspace}

\title{Path-length dependence of parton and jet energy loss from universal scaling laws}

\author*[a]{François Arleo}
\author[b]{Guillaume Falmagne}

\affiliation[a]{SUBATECH UMR 6457 (IMT Atlantique, Universit\'e de Nantes, IN2P3/CNRS)\\ 4 rue Alfred Kastler, 44307 Nantes, France}

\affiliation[b]{High Meadows Environmental Institute, Guyot Hall, Princeton University,\\ Princeton, NJ 08544-1003, USA}

\emailAdd{francois.arleo@cern.ch}
\emailAdd{g.falmagne@princeton.edu}

\abstract{The universal dependence of hadron suppression, $R_{\rm{AA}}(p_\perp)$, observed at large-$p_\perp$ in heavy ion collisions at RHIC and LHC allows for a systematic determination of the average parton energy loss $\langle \epsilon \rangle$ in quark-gluon plasma (QGP). A simple relation between $\langle \epsilon \rangle$ and the soft particle multiplicity allows for probing the dependence of parton energy loss on the medium path-length. We find that all the available measurements are consistent with $\langle \epsilon \rangle \propto L^\beta$ with $\beta=1.02\pm^{0.09}_{0.06}$, consistent with the pQCD expectation of parton energy loss in a longitudinally expanding QGP. We then show, based on the model predictions, that the data on the azimuthal anisotropy coefficient divided by the collision eccentricity, $v_2/\rm{e}$, follows the same scaling property as $R_{\rm{AA}}$. Finally, a linear relationship between $v_2/\rm{e}$ and the logarithmic derivative of $R_{\rm{AA}}$ at large $p_\perp$ offers a purely data-driven access to the $L$ dependence of parton energy loss. Quite remarkably, both hadron and jet measurements obey this latter relationship, moreover with consistent values of $\beta$. This points to the same parametric path-length dependence of parton and jet energy loss in QGP.}

\FullConference{42nd International Conference on High Energy Physics (ICHEP2024)\\
18-24 July 2024\\
Prague, Czech Republic\\}

\begin{document}
\maketitle

The theory of parton energy loss in quark-gluon plasma (QGP) and its associated jet quenching phenomenology in heavy ion collisions has evolved significantly over the past decade, driven by precise measurements from RHIC and LHC. Despite these advancements, fundamental questions remain, such as how parton energy loss in QGP depends on the medium path-length $L$. In these proceedings, we discuss how this question can be addressed through a data-driven strategy. More specifically, we will exploit the scaling properties of large-$\pt$ hadron spectra in heavy ion collisions, predicted by a simple analytic energy loss model and observed in experimental data. In addition, we show that \textit{jet measurements} obey one such scaling law, offering the possibility to study jet energy loss in QGP within the same strategy.

\begin{figure}[b]
        \centering
        \includegraphics[width=\linewidth]{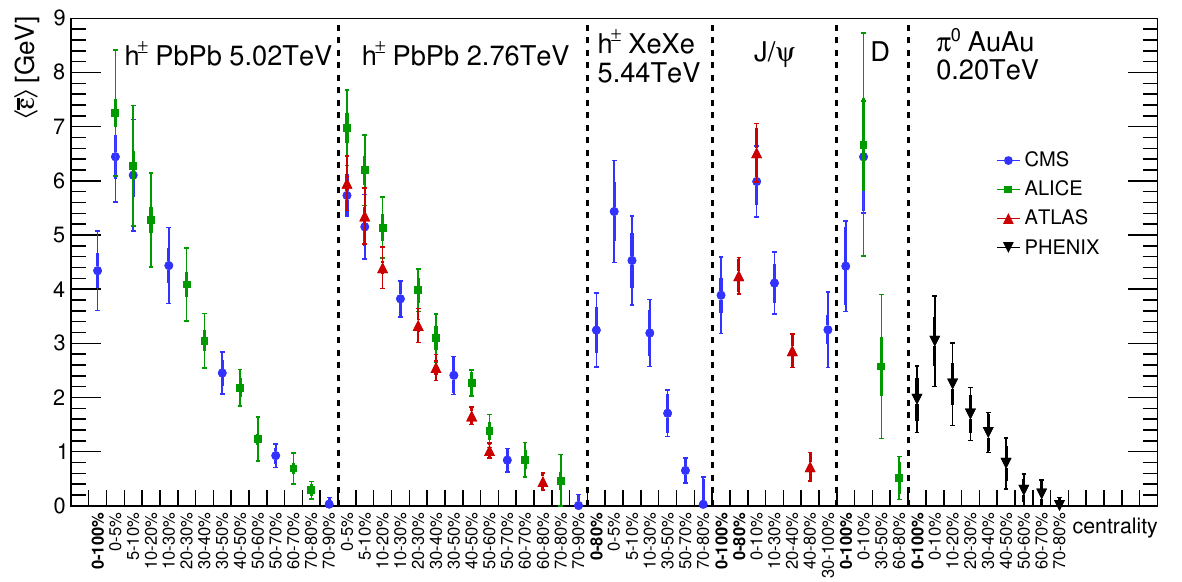} 
        \caption{Energy loss scales  $\meanepsbar$ extracted from light ($\pi^0$, $h^\pm$) and heavy ($\jpsi$, $D$
        in PbPb at 5.02\TeV
        ) hadron data in various collision systems and centralities.
        }
        \label{fig:allscales}
\end{figure}

As discussed in Refs.~\cite{Arleo:2017ntr,Arleo:2022shs}, the quenching of hadron spectra,
$\RAA(\pt)$, exhibits at large $\pt$ a universal shape as a function of $\pt$: 
\be
\label{eq:scaling1}
\RAA^{h}(\pt, \meanepsbar, n) \simeq f(u\equiv \pt/\, n\meanepsbar). 
\ee
Here, $n$ represents the spectral index of the hadron \pt spectrum in pp collisions, $\dsigpp^h/\dd\pt \propto \pt^{-n}$, and $\meanepsbar$ denotes a characteristic energy loss scale reflecting the properties of the medium produced in a given collision system. Fits to the quenching of hadron spectra, measured in heavy ion collisions at RHIC (AuAu collisions) and LHC (XeXe, PbPb), allow for a systematic extraction of the scales $\meanepsbar$ across a broad range of collision energies $\sqrts$ and centrality classes ${\cal C}$, as illustrated in Fig.~\ref{fig:allscales} (see Ref.~\cite{Arleo:2022shs} for more details and for the bibliographical references therein to the data sets). 

In order to go a step further, it is instructive to relate the $\meanepsbar$ to the relevant physical quantities in heavy ion collisions. In the BDMPS formalism, $\meanepsbar$ should scale according to~\cite{Arleo:2022shs}
\be
\label{eq:mean_eloss_expanding}
    \meanepsbar \propto \langle z \rangle\, C_k\, \qzero\, \tau_0^\alpha\,L^{2-\alpha}\,,
\ee 
where $\langle z \rangle$ is the momentum fraction carried by the hadron, $C_k$ is the color charge of the parton $k$, $L$ is the medium path-length and $\qzero$ stands for the transport coefficient at formation time $\tau_0$ ($\tau_0 \ll L$). Since the values of  $\langle z \rangle$ and $C_k$ may differ for heavy hadrons, the analysis will focus on light hadrons from now on. The parameter $\alpha$ entering \eqref{eq:mean_eloss_expanding} characterizes the dynamical expansion of the medium, according to $\qhat(\tau) = \qhat_0 \left(\tau_0/\tau\right)^{\alpha}$. 
The initial transport coefficient \mbox{$\qzero = \qhat(\tau_0)$} is proportional to the initial parton density $n_0$, which in the Bjorken picture can be estimated as
\be\label{eq:densitydNch}
n_0 = \frac{1}{\AT \tau_0}\,\left.\dNg\right|_{y=0} \,\simeq\,\frac{3}{2}\,\frac{1}{\AT \tau_0}\,\left.\dNch\right|_{y=0} \, ,
\ee
where $\AT$ is the transverse overlap area of the two crossing nuclei. Plugging \eqref{eq:densitydNch} in \eqref{eq:mean_eloss_expanding} leads to the following parametric dependence,~\cite{Arleo:2022shs}
\be\label{eq:eloss_scaling}
\meanepsbar\, = K \times \, \left(\frac{1}{\AT}\,\dNch \,L^{\beta}\right) \,,
\ee
with $\beta=2-\alpha$. The geometric quantities $A_\perp$ and $L$ entering Eq.~\eqref{eq:eloss_scaling} are determined through an optical Glauber model, assuming hard sphere nuclear densities. The charged particle multiplicity data $\dd N_\textnormal{ch}/\dd \eta$ at mid-rapidity is taken from PHENIX at RHIC and from \mbox{ALICE} and CMS at LHC (see~\cite{Arleo:2022shs} and references therein). 
The energy loss scales $\meanepsbar$ extracted from the quenching of light hadrons are fitted using Eq.~\eqref{eq:eloss_scaling}, with $\beta$ (and $K$) taken as a free parameter.

\begin{figure}[ht]
    \centering
    \begin{minipage}{0.5\textwidth}
        \centering
        \includegraphics[width=\linewidth]{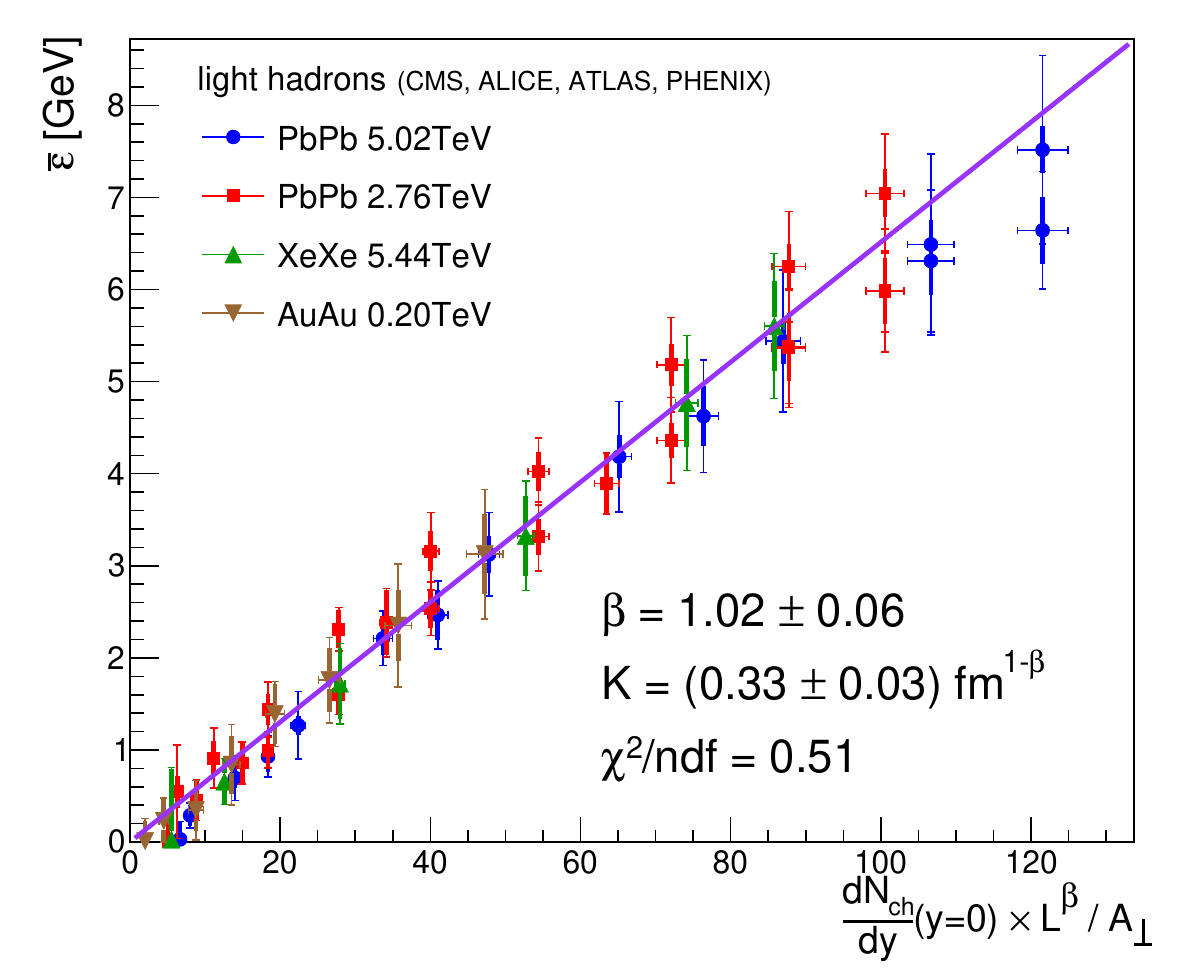}
        \caption{Energy loss scales $\meaneps$ in various collision systems as a function of $\dd N_\textnormal{ch}/\dd y\times L^{\beta}/\AT$.}
        \label{fig:scaling2}
    \end{minipage}
    \hfill
    \begin{minipage}{0.42\textwidth}
        \centering
        \includegraphics[width=\linewidth]{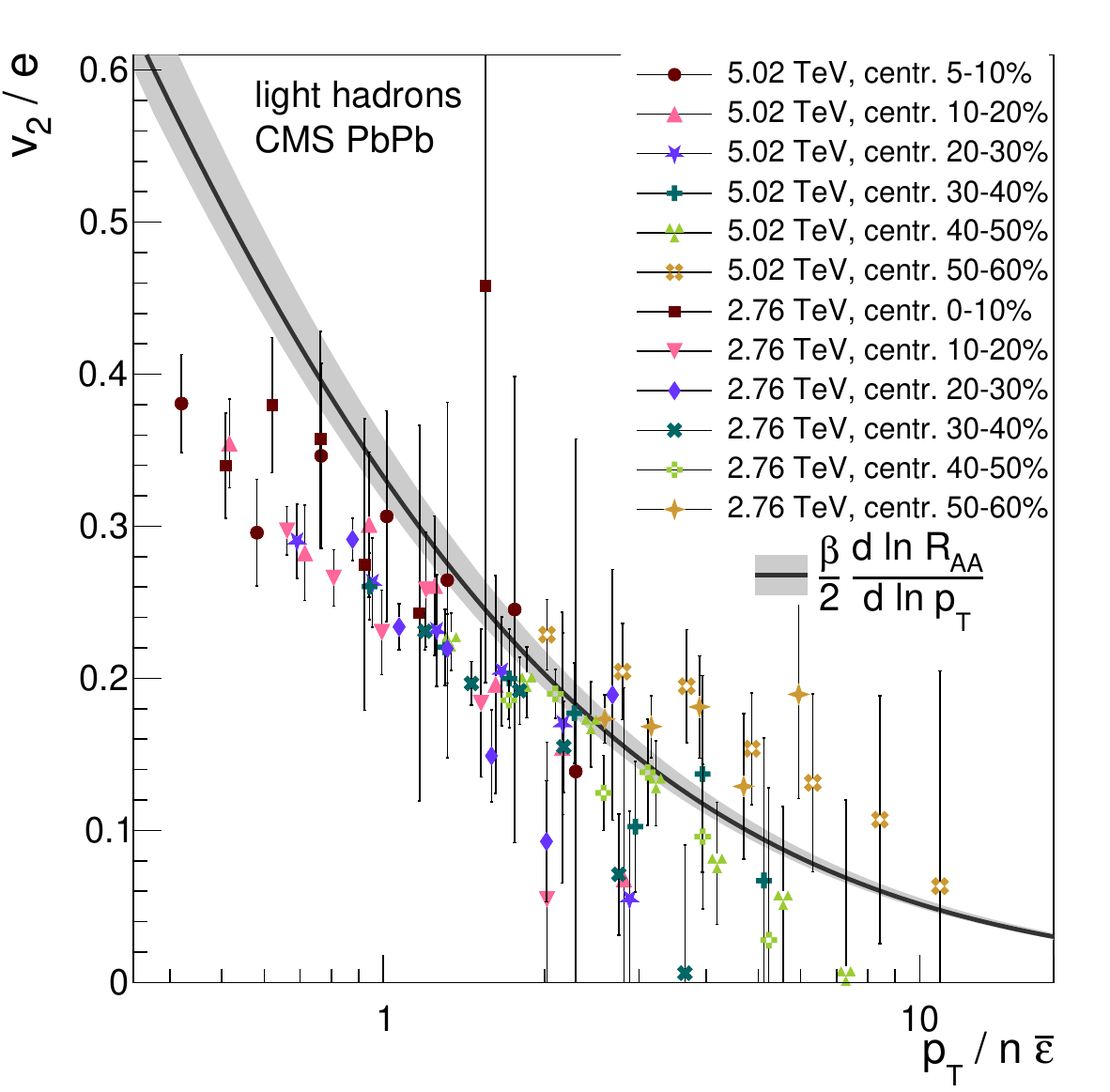}
        \caption{$v_2/\ecc$ of light hadrons vs. $\pt/n\meanepsbar$. The band corresponds to $\beta=1.02^{+0.09}_{-0.06}$.}
        \label{fig:v2scaling}
    \end{minipage}
\end{figure}

As can be seen in Figure~\ref{fig:scaling2}, the scaling property expected from \eq{eq:eloss_scaling} holds very well for light hadron production in all collision systems. The fits leads to $\beta=1.02{\raisebox{0.5ex}{\tiny$\substack{+0.09\\-0.06}$}}
$ 
($\chi^2/\textnormal{ndf}=0.51$), where the uncertainties originate from the fit and from the use of alternative Glauber models~\cite{Arleo:2022shs}. The value of $\beta$ proves compatible with unity, that is the BDMPS expectation in QGP experiencing a purely longitudinal expansion (i.e. $\alpha=1$).

\begin{figure}[ht]
    \centering    
    \includegraphics[width=0.48\textwidth]{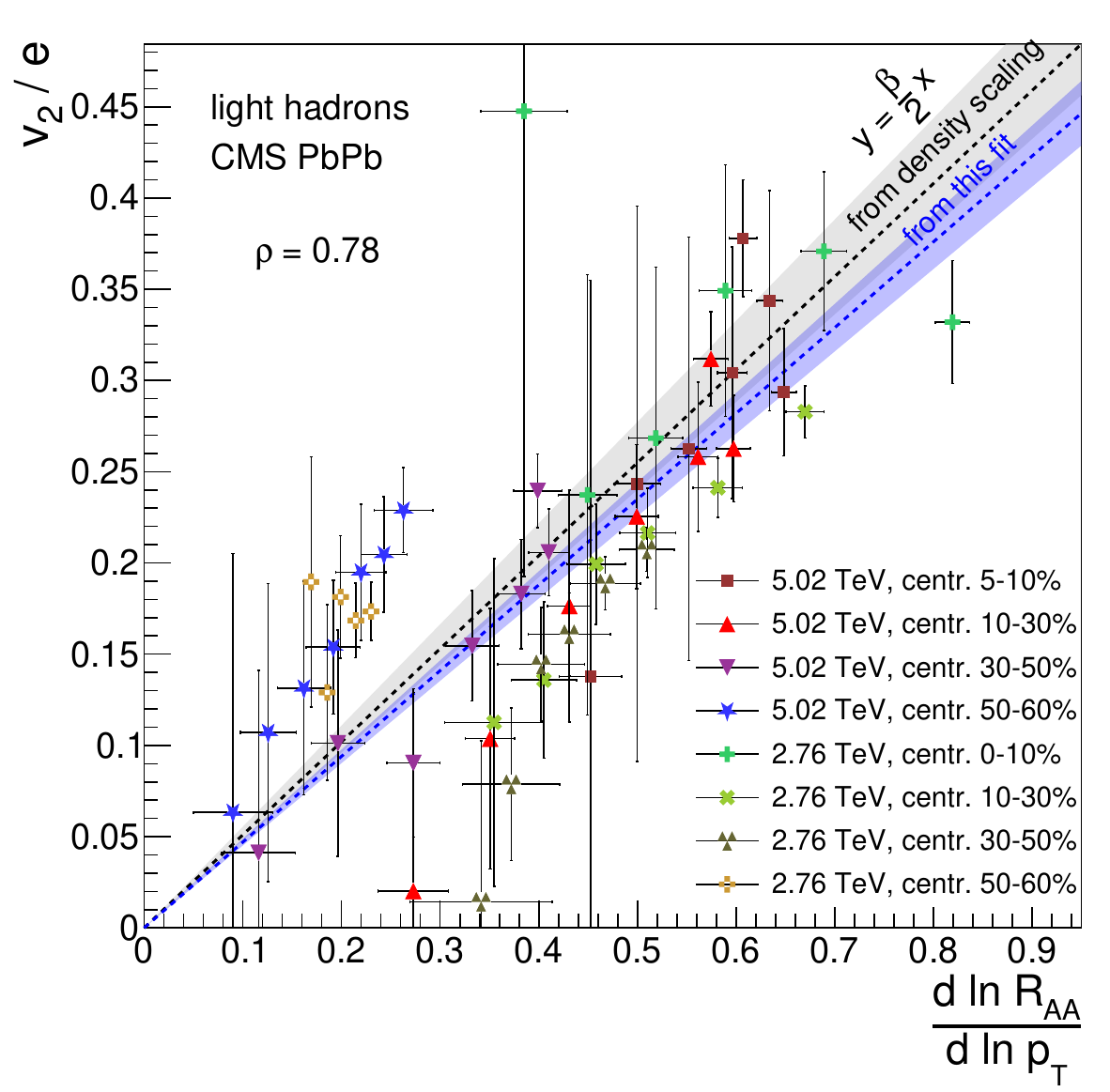}
    \hfill
    \includegraphics[width=0.48\textwidth]{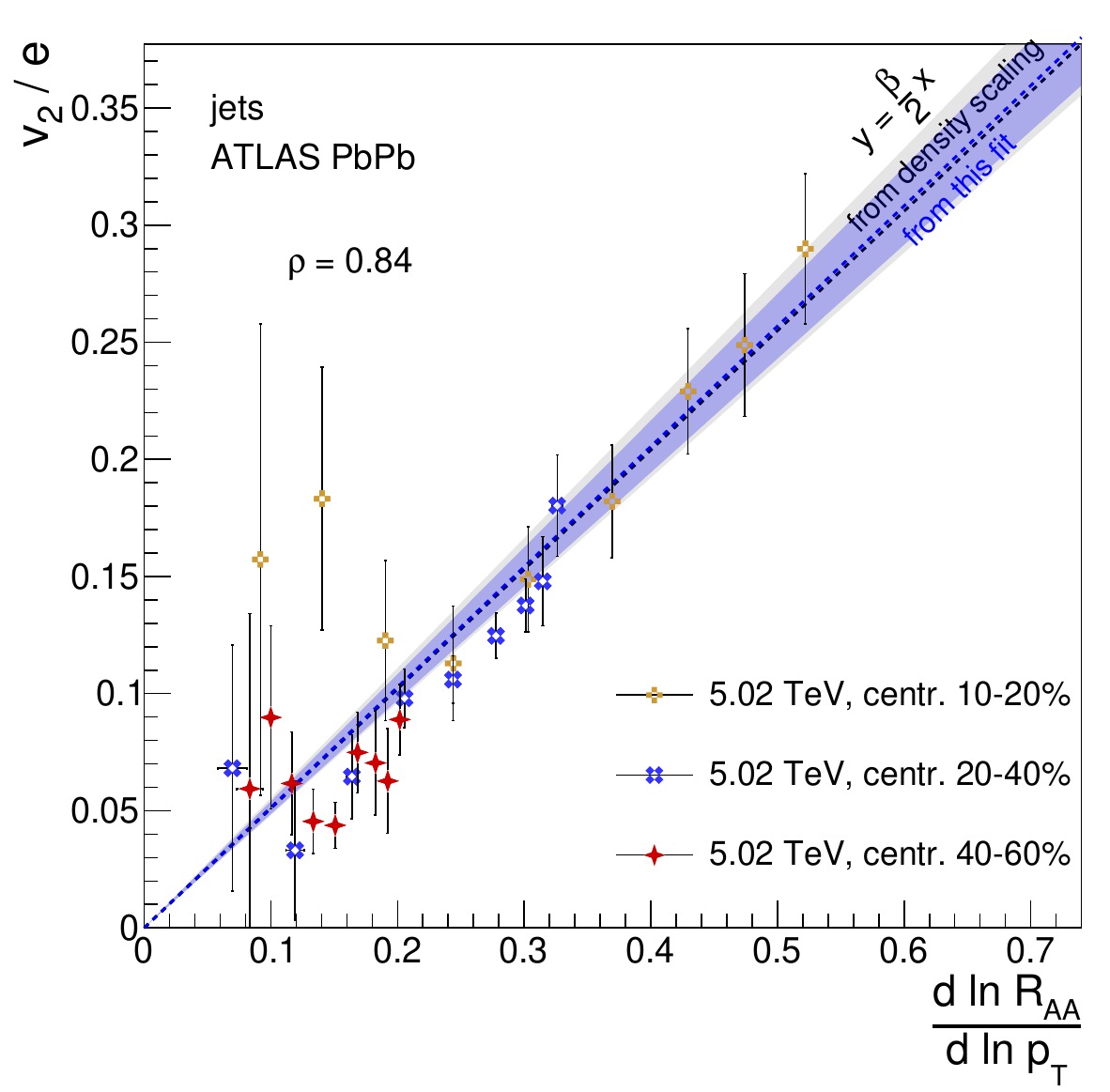}
    \caption{Relation between $v_2/\ecc$ and $\dd\ln\RAA/\dd\ln\pt$ for light hadrons
    in CMS~\cite{CMS:2012tqw,CMS:2017xgk,CMS:2012aa,CMS:2016xef} (left) and 
    inclusive jets in ATLAS~\cite{ATLAS:2018gwx,ATLAS:2021ktw} (right). 
    The bands \mbox{$y=\beta\,x/2$} show the expectation from~\eqref{eq:v2finalb}, with the value \mbox{$\beta=1.02^{+0.09}_{-0.06}$} determined from the density scaling fit of hadrons of Eq.~\eqref{eq:eloss_scaling} (grey band, left), and the $\beta$ values extracted from linear fits of the displayed hadron or jet data (blue bands).
    }
    \label{fig:v2vsRAA}
\end{figure}

Once the dependence of $\meanepsbar$ with $L$ is empirically determined, it becomes possible to investigate the azimuthal dependence of hadron suppression, from which the $v_2$ coefficient can be computed. Using the universal shape of $\RAA$, its $\phi$ dependence can be modeled as
\be\label{eq:raa_phi}
\RAA(u, n, \phi) = f\left(u\times \left(L/L(\phi)\right)^\beta, n\right)\,,
\ee
where $L(\phi)$ is the path-length as a function of the azimuthal angle $\phi$ (with $\phi=0$ being the direction of the impact parameter), approximated as 
$L(\phi) = L\times\left(1-\ecc\, \cos\left(2\phi\right)\right)$, where the eccentricity is defined as $\ecc = \left( L(\pi/2)-L(0)\right) / \left( L(\pi/2)+L(0) \right)$.
To first order in $\ecc$ and using Eq.~\eqref{eq:raa_phi}, the elliptic flow coefficient can be approximated as~\cite{Arleo:2022shs}
\bea
\frac{v_2(u, n)}{\ecc} &\simeq& \frac\beta{2}\,\frac{\partial \ln f(u, n)}{\partial \ln u}\,, \label{eq:v2finala}\\
\frac{v_2(\pt)}{\ecc}&\simeq& \frac{\beta}{2}\,\frac{\pt}{\RAA(\pt)}\,\frac{\partial \RAA(\pt)}{\partial \pt}\,.\label{eq:v2finalb}
\eea
Eq.~\eqref{eq:v2finala}  indicates that $v_2/\ecc$ at large $\pt$ should exhibit the same approximate universal dependence on $\pt/n\meanepsbar$ as $\RAA$, for all collision energies and centrality classes. In order to check this prediction, $\RAA(\pt, \phi)$ has been computed in each collision system, using our full model including Glauber simulations; this is then fitted to extract $v_2$ values. The model predictions using the $\beta$ value fitted in Fig.~\ref{fig:scaling2} (grey band) and the CMS measurements of $v_2$ in \PbPb collisions~\cite{CMS:2012tqw,CMS:2017xgk} are plotted together in Fig.~\ref{fig:v2scaling} as a function of the scaling variable $\pt/n\meanepsbar$, where the values of $\meanepsbar$ are determined from the fits of $\RAA$ (see Fig.~\ref{fig:allscales}). The predicted scaling for the different collision systems is clearly apparent in the data, except at the lowest values of $\pt/n\meanepsbar$.
The simultaneous calculation of $\RAA$ and $v_2$ in the model and its agreement with measurements of both observables is thus compelling~--~moreover with a path-length dependence compatible with perturbative QCD.
In addition, Eq.~\eqref{eq:v2finalb} indicates that $v_2$ and $\RAA$ at a given $\pt$ are trivially related for measurements from the same collision system. In particular, this relation does not involve the knowledge of the energy loss scale $\meanepsbar$ nor that of the function $f$ appearing in~\eqref{eq:v2finala}. It is also worth noting that the normalization uncertainties of $\RAA$ vanish when computing~\eqref{eq:v2finalb}.
We would like to check whether it also holds when comparing solely $\RAA$ and $v_2$ \emph{measurements}, independently of the present energy loss model. The $v_2/\ecc$ measurements are plotted as a function of the slope $\dd\ln\RAA/\dd\ln\pt$ in Fig.~\ref{fig:v2vsRAA} (left). The slope is evaluated with an agnostic fit of CMS $\RAA$ data~\cite{CMS:2012aa,CMS:2016xef}; when the centrality classes for $\RAA$ and $v_2$ did not match, values for combined classes have been interpolated with weights scaling with $N_{\textrm{coll}}$.

Although the present precision of the data does not allow yet for a rigorous test of Eq.~\eqref{eq:v2finalb}, the correlation between the two measured quantities is clearly apparent (correlation coefficient $\rho=0.78$). 
In addition, a fit with the expected function $y=\beta\,x/2$ reproduces fairly the hadron observations, resulting in a value $\beta = 0.94\pm0.04$ (fit uncertainties only) consistent with the value $\beta=1.02{\raisebox{0.5ex}{\tiny$\substack{+0.09\\-0.06}$}}
$ 
extracted from the scaling with hadron multiplicity, Eq.~\eqref{eq:eloss_scaling}. This gives confidence that the relation between $\RAA$ and $v_2$ at large $\pt$ indeed provides a direct experimental access to the path-length dependence of parton energy loss in QGP. 
A possible explanation for the slight overshoot in the $50$--$60\%$ centrality class may be back-to-back jet correlations contamination in data~\cite{CMS:2016xef}. In other centrality classes, disagreement may be due to the overestimation of \ecc in the Glauber model presently used, which relies on hard sphere nuclear densities.

Although related in principle, hadron and jet measurements in heavy ion collisions are likely to probe different aspects of medium-induced gluon radiation. The former appears as a good proxy of genuine \emph{parton} energy loss in QGP, the latter probes instead gluon emission off a final state made of multiple particles acting coherently. Consequently, there is no reason a priori that parton and jet energy loss exhibit the same energy and path-length parametric dependence, nor the $\pt$ dependence of $\RAA$ of both processes to be identical.\footnote{The $\RAA$ and $v_2$ of jets have also been discussed at this conference, see Ref.~\cite{Mehtar-Tani:2024jtd,Ogrodnik:2024qug} for details.} 
It is therefore worthwhile to investigate whether the relation~\eqref{eq:v2finalb} between $\RAA$ and $v_2$, derived on rather general grounds, also applies to jet production --~despite the different $\pt$-dependence of $\RAA$~-- and, if so, to determine which value of $\beta$ best matches the data.

The results, using $\RAA$~\cite{ATLAS:2018gwx} and $v_2$~\cite{ATLAS:2021ktw} ATLAS data in PbPb collisions at $\sqrts=5.02$~TeV, are shown in Fig.~\ref{fig:v2vsRAA} (right). As for hadrons, the expected function $y=\beta\,x/2$ fits well the data\footnote{The centrality class $0$--$5\%$ for hadrons and $0$--$10\%$ for jets have not been included as the small eccentricity results in large uncertainties on $v_2/\ecc$; in addition, the eccentricity in central collisions depends on density fluctuations that cannot be accounted for in the optical Glauber model we use to compute it.}, with a value $\beta_{\textrm{jets}} = 1.03\pm0.06$ which is consistent with the value extracted from Eq.~\eqref{eq:eloss_scaling} for light hadrons. This is a remarkable indication that parton and jet energy loss share the same parametric, close-to-linear, dependence on the medium path-length. 
It is also interesting to note that a different path-length dependence of \textit{jet} energy loss, $\Delta E \propto L^{0.59}$, is obtained in the data-driven approach of Ref.~\cite{Wu:2023azi}, while other studies point instead towards a stronger dependence on $L$, $\Delta E \propto L^{1.4}$~\cite{Djordjevic:2018ita} and $\Delta E \propto L^{2}$~\cite{Ogrodnik:2024qug}.

In summary, relating the values of $\meanepsbar$ extracted from $\RAA$ to the hadron multiplicity enables the determination of the $L$ dependence of parton energy loss, $\meanepsbar \propto L^{\beta}$ with $\beta=1.02{\raisebox{0.5ex}{\tiny$\substack{+0.09\\-0.06}$}}
$, in agreement with a longitudinally expanding QGP. The $v_2/\ecc$ anisotropy coefficient exhibits the same scaling property as $\RAA$, in both the model and data. Finally we highlight a relation between $v_2/\ecc$ and $\RAA$, predicted in the model and observed in data, which allows for a data-driven extraction of the path-length dependence of parton energy loss. We find that jet data obey the same relation, pointing to a similar path-length dependence of parton and jet energy loss in quark-gluon plasma. The LHC Runs~3 \& 4 should allow for testing with unprecedented precision these  scaling properties.

\acknowledgments
GF acknowledges the support from William Miller. This work is funded by the ``Agence Nationale de la Recherche'' under grant ANR-18-CE31-0024-02.

\providecommand{\href}[2]{#2}\begingroup\raggedright\endgroup


\begin{thebibliography}{10}

\bibitem{Arleo:2017ntr}
F.~Arleo, \emph{{Quenching of Hadron Spectra in Heavy Ion Collisions at the LHC}}, \href{https://doi.org/10.1103/PhysRevLett.119.062302}{\emph{Phys. Rev. Lett.} {\bfseries 119} (2017) 062302} [\href{https://arxiv.org/abs/1703.10852}{{\ttfamily 1703.10852}}].

\bibitem{Arleo:2022shs}
F.~Arleo and G.~Falmagne, \emph{{Probing the path-length dependence of parton energy loss via scaling properties in heavy ion collisions}}, \href{https://doi.org/10.1103/PhysRevD.109.L051503}{\emph{Phys. Rev. D} {\bfseries 109} (2024) L051503} [\href{https://arxiv.org/abs/2212.01324}{{\ttfamily 2212.01324}}].

\bibitem{CMS:2012tqw}
{\scshape CMS} collaboration, \emph{{Azimuthal anisotropy of charged particles at high transverse momenta in PbPb collisions at $\sqrt{s_{NN}}=2.76$ TeV}}, \href{https://doi.org/10.1103/PhysRevLett.109.022301}{\emph{Phys. Rev. Lett.} {\bfseries 109} (2012) 022301} [\href{https://arxiv.org/abs/1204.1850}{{\ttfamily 1204.1850}}].

\bibitem{CMS:2017xgk}
{\scshape CMS} collaboration, \emph{{Azimuthal anisotropy of charged particles with transverse momentum up to 100 GeV/ c in PbPb collisions at $\sqrt {s}_{{NN}}$=5.02 TeV}}, \href{https://doi.org/10.1016/j.physletb.2017.11.041}{\emph{Phys. Lett. B} {\bfseries 776} (2018) 195} [\href{https://arxiv.org/abs/1702.00630}{{\ttfamily 1702.00630}}].

\bibitem{CMS:2012aa}
{\scshape CMS} collaboration, \emph{{Study of high-$p_\perp$ charged particle suppression in PbPb compared to pp collisions at $\sqrt{s_{\mathrm{NN}}}=$2.76 TeV}}, \href{https://doi.org/10.1140/epjc/s10052-012-1945-x}{\emph{Eur.\ Phys.\ J.} {\bfseries C72} (2012) 1945} [\href{https://arxiv.org/abs/1202.2554}{{\ttfamily 1202.2554}}].

\bibitem{CMS:2016xef}
{\scshape CMS} collaboration, \emph{{Charged-particle nuclear modification factors in PbPb and pPb collisions at $ \sqrt{s_{\mathrm{N}\;\mathrm{N}}}=5.02 $ TeV}}, \href{https://doi.org/10.1007/JHEP04(2017)039}{\emph{JHEP} {\bfseries 04} (2017) 039} [\href{https://arxiv.org/abs/1611.01664}{{\ttfamily 1611.01664}}].

\bibitem{ATLAS:2018gwx}
{\scshape ATLAS} collaboration, \emph{{Measurement of the nuclear modification factor for inclusive jets in Pb+Pb collisions at $\sqrt{s_\mathrm{NN}}=5.02$ TeV with the ATLAS detector}}, \href{https://doi.org/10.1016/j.physletb.2018.10.076}{\emph{Phys. Lett. B} {\bfseries 790} (2019) 108} [\href{https://arxiv.org/abs/1805.05635}{{\ttfamily 1805.05635}}].

\bibitem{ATLAS:2021ktw}
{\scshape ATLAS} collaboration, \emph{{Measurements of azimuthal anisotropies of jet production in Pb+Pb collisions at $\sqrt{s_{NN}} =$ 5.02 TeV with the ATLAS detector}}, \href{https://doi.org/10.1103/PhysRevC.105.064903}{\emph{Phys. Rev. C} {\bfseries 105} (2022) 064903} [\href{https://arxiv.org/abs/2111.06606}{{\ttfamily 2111.06606}}].

\bibitem{Mehtar-Tani:2024jtd}
Y.~Mehtar-Tani, D.~Pablos and K.~Tywoniuk, \emph{{Jet suppression and azimuthal anisotropy from RHIC to LHC}}, \href{https://doi.org/10.1103/PhysRevD.110.014009}{\emph{Phys. Rev. D} {\bfseries 110} (2024) 014009} [\href{https://arxiv.org/abs/2402.07869}{{\ttfamily 2402.07869}}] and these proceedings.

\bibitem{Ogrodnik:2024qug}
A.~Ogrodnik, M.~Ryb\'a\v{r} and M.~Spousta, \emph{{Flavor and path-length dependence of jet quenching from inclusive jet and \ensuremath{\gamma}-jet suppression}},  \href{https://arxiv.org/abs/2407.11234}{{\ttfamily 2407.11234}} and these proceedings.

\bibitem{Wu:2023azi}
J.~Wu, W.~Ke and X.-N.~Wang, \emph{{Bayesian inference of the path-length dependence of jet energy loss}}, \href{https://doi.org/10.1103/PhysRevC.108.034911}{\emph{Phys. Rev. C} {\bfseries 108} (2023) 034911} [\href{https://arxiv.org/abs/2304.06339}{{\ttfamily 2304.06339}}].

\bibitem{Djordjevic:2018ita}
M.~Djordjevic, D.~Zigic, M.~Djordjevic and J.~Auvinen, \emph{{How to test path-length dependence in energy loss mechanisms: analysis leading to a new observable}}, \href{https://doi.org/10.1103/PhysRevC.99.061902}{\emph{Phys. Rev. C} {\bfseries 99} (2019) 061902} [\href{https://arxiv.org/abs/1805.04030}{{\ttfamily 1805.04030}}].

\end{thebibliography}
\end{document}